\documentclass[12pt]{article}%
\usepackage{amsmath}
\usepackage{amsfonts}
\usepackage{amssymb}
\usepackage{graphicx}
\usepackage[T1]{fontenc}
\usepackage[utf8]{inputenc}
\usepackage{authblk}
\usepackage{setspace}
\usepackage[margin=1.3in]{geometry}
\begin{document}

\title{Statistical harmonization and uncertainty assessment in the comparison of
satellite and radiosonde climate variables}
\author[1]{F. Finazzi \thanks{Contact author: francesco.finazzi@unibg.it}}
\author[1]{A. Fass\`{o}}
\author[2]{ F. Madonna}
\author[1]{I. Negri}
\author[3]{B. Sun}
\author[2]{M. Rosoldi}

\affil[1]{University of Bergamo, Department of Management, Information and
Production Engineering. Italy}
\affil[2]{National Research Council of Italy. Institute of Methodologies for Environmental Analysis}
\affil[3]{NOAA Center for Satellite Applications and Research, College Park, Maryland. USA}

\renewcommand\Authands{ and }
\maketitle

\doublespacing

\begin{abstract}
Satellite product validation is key to ensure the delivery of quality products
for climate and weather applications. To do this, a fundamental step is the
comparison with other instruments, such us radiosonde. This is specially true
for Essential Climate Variables such as temperature and humidity.

Thanks to a functional data representation, this paper uses a likelihood based
approach which exploits the measurement uncertainties in a natural way. In
particular the comparison of temperature and humdity radiosonde measurements
collected within RAOB network and the corresponding atmospheric profiles
derived from IASI interferometers aboard of Metop-A and Metop-B satellites is
developed with the aim of understanding the vertical smoothing mismatch uncertainty.

Moreover, conventional RAOB functional data representation is assessed by
means of a comparison with radiosonde reference measurements given by GRUAN
network, which provides high resolution fully traceable radiosouding profiles.
In this way the uncertainty related to coarse vertical resolution, or
sparseness, of conventional RAOB is assessed.

It has been found that the uncertainty of vertical smoothing mismatch averaged
along the profile is $0.50$ $K$ for temperature and $0.16$ g/kg for water
vapour mixing ratio. Moreover the uncertainty related to RAOB sparseness,
averaged along the profile is $0.29$ $K$ for temperature and $0.13$ g/kg for
water vapour mixing ratio.

\end{abstract}


\section{Introduction}

Satellite validation is key to ensure that satellite products meet the mission
specified requirements for climate and weather applications. Since the
agreement of satellite measurements with ground-based reference measurements
is an essential quality indicator, one major issue in performing a
rigorous validation is the quantification of the uncertainty due to the
co-location mismatch in time and space between satellite and ground based
reference observations. This mismatch is due to the different sampling of
atmosphere carried out by the two instruments (Verhoelst et al., 2015), which
are also quite often based on very different sensing techniques.
Moreover, satellite and ground based observations are typically collected on
very different time scales and spatial scales. As a consequence, in a
satellite vs ground measurements comparison, we may have horizontal and/or
vertical and/or temporal mismatches in smoothing and/or profile resolution,
which contribute to the overall co-location mismatch uncertainty (hereinafter
mismatch uncertainty).

Over the last decade, several authors tried to estimate the impact of the
satellite vs. ground based mismatch uncertainty of the
Essential Climate Variables (ECVs). The most common approach is to use simple
descriptive statistics to identify the maximum temporal and spatial distances
which warranty a controlled mismatch uncertainty. See e.g. Pappalardo et al.
(2010) for aerosols and Kursinski \& Hajj (2001) for water vapor. Moreover,
considering vertical resolution, Pougatchev et al. (2007) found that, when
available, the averaging kernels can be used to reconcile the vertical
resolution of satellite and ground based observations of temperature and humidity.

A rigorous metrological characterization of the mismatch uncertainty requires
the quantification of the total uncertainty budget for each satellite
retrieved ECV. Hence, the uncertainty budget includes the contribution of
random, systematic, sampling, smoothing uncertainties and their correlation
with all the relevant environmental factors. Pioneering works in this
direction are Ridolfi et al. $\left(  2007\right)  $ and Lambert et al.
$\left(  2011\right)  $. More recently, Verhoelst et al. (2015) used an
explicit physic simulation method for computing a full uncertainty budget
closure for ozone. Moreover, Fass\`{o} et al. (2014) and Ignaccolo et al.
$\left(  2015\right)  $ proposed an approach based on the extension of the
classical functional regression model able to cover for hetereoskedasticity of
mismatch error in temperature and humidity observation.

GAIA-CLIM project (www.gaia-clim.eu) is a Horizon 2020 project which aims at
improving the use of non-satellite measurements to characterise, calibrate and
validate satellite measurements. Considering temperature and humidity, one of
its objectives is to understand the mismatch uncertainty in the comparison of
the satellite observations obtained by the Infrared Atmospheric Sounding
Interferometer (IASI) instrument, on board EUMETSAT MetOP-A and --B, with the
radiosonde observations (RAOBs). In fact RAOB profiles are appealing for
satellite validation because of their extensive spatial and temporal coverage,
hence permitting the assessment of mismatch uncertainty at global level.
Despite of this, RAOB observations cannot be strictly considered reference
measurements because they are not fully traceable and have a limited vertical
resolution, see Dirksen et al. $\left(  2014\right)  $. Note that the GCOS
Reference Upper-Air Network (GRUAN, www.gruan.org) provides reference
products, which are fully traceable but have a very limited spatial coverage
(16 sites as the paper is written).

Along these lines, the present paper focuses on the vertical smoothing
mismatch uncertainty of IASI-RAOB profile comparison of temperature and
humidity. This objective is achieved by a statistical technique for vertical
harmonization which is independent on the availability of IASI averaging
kernels, hence especially relevant for comparisons where averaging kernels are
not available, in particular for historical data analysis. To do this, the
vertical data-point sparseness of RAOB network is assessed by means of a
comparisons with GRUAN reference products where available.

The proposed technique is a two steps technique. At the first step RAOB
profiles are transformed into continuous functions using splines, which are
optimized to match as close as possible to GRUAN profiles. In doing this,
vertical sparseness uncertainty and processing mismatch uncertainty are
assessed. At the second step RAOB profiles are harmonized by considering
weighting functions based on the Generalized Extreme Values (GEV) probability
density function (pdf) whose parameters depend on the IASI levels.

The paper is structured as follows. In Section \ref{sec:datasets} data from
both satellite (IASI) and radiosonde (GRUAN and RAOB) are introduced. In
Section \ref{sec:mismatches}, the various sources of uncertainty arising in
satellite-ground comparisons are reviewed. Sections
\ref{sec:stat.harmonization} and \ref{sec:likelihood} discuss novel
statistical modeling: the former section leverages on intuition while the
latter embeds the same model in a rigorous maximum likelihood estimation
problem. Section \ref{sec:Case.Study} applies this approach to IASI-RAOB
comparison for a number of RAOB stations in central Europe. To do this, in
Section \ref{sec:Case.Study} the RAOB soundings are transformed into
functional data and harmonized to match IASI vertical smoothing. Then the
sparseness uncertainty of RAOB and vertical smoothing uncertainty of IASI-RAOB
comparison are computed. Section \ref{sec:Conclusions} gives concluding remarks.

\section{Data sets}\label{sec:datasets}

\begin{figure}[ptb]
\begin{center}
\includegraphics[width=4in]{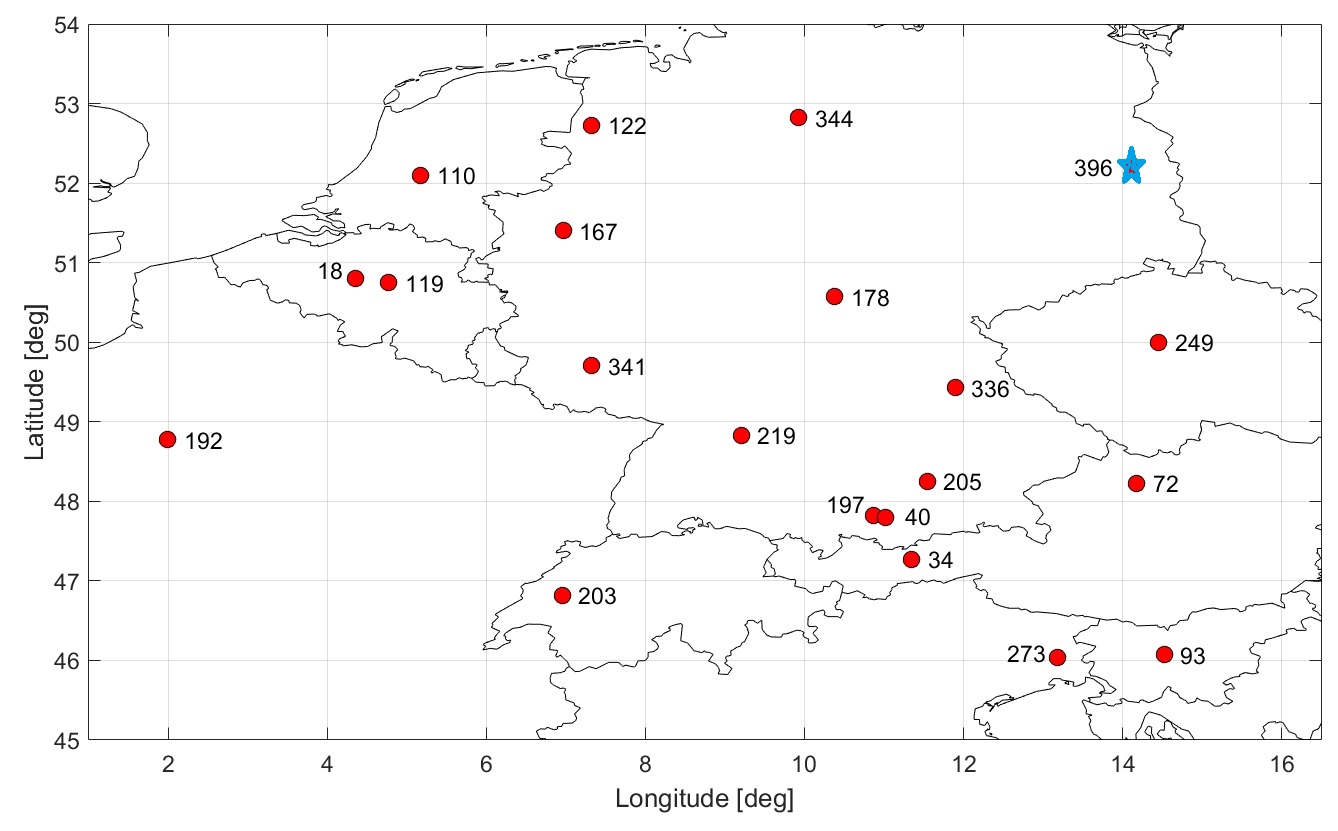}
\end{center}
\caption{Spatial distribution of RAOB stations (red circles) in Central Europe
and RAOB-GRUAN station in Lindenberg (cyan star). The number near each dot is
the number of launches during the period January 2015 - February 2016. }%
\label{fig:station_map}%
\end{figure}

The data sets used in this study include atmospheric profile retrievals
derived from IASI instrument and from conventional (RAOB) and reference
(GRUAN) radiosonde networks. The RAOB-IASI co-location data set, which is
provided by NOAA-NESDIS, has been collected through the NOAA Products
Validation System (NPROVS), see Reale et al. $\left(  2010\right)  $ and Sun
et al. $\left(  2017\right)  ,$ (http://www.star.%
%TCIMACRO{\TeXButton{\-}{\-}}%
%BeginExpansion
\-%
%EndExpansion
nesdis.noaa.gov/smcd/opdb/nprovs/). The NPROVS data set used in this study
includes $K=3908$ co-located profiles at 21 RAOB stations selected across the
central European area (C-EU), described in Figure \ref{fig:station_map}, for
the period January 2015 -- February 2016.

Each co-location pair includes RAOB and IASI profiles for temperature and
water vapor mixing ratio (WVMR) with RAOB at mandatory and significant levels
and IASI at 100 levels. In order to consider profiles with enough data, the
data set has been filtered as follows: temperature (WVMR) has been restricted
to atmospheric range $958.6-10$ $hPa$ ($958.6-300$ $hPa$) and only
co-locations with at least $20$ $\left(  14\right)  $ RAOB measurements have
been selected for the analysis, giving $K=1596$ ($2648$) co-locations out of
the original $K=3908$ NPROVS co-locations. Notice that the atmospheric range
considered after filtering is still relevant for climate and weather studies
as pressure levels $10$ $hPa$ and $300$ $hPa$, correspond to around $40$ $km$
and $10$ $km$ respectively.

In addition, the GRUAN station at Lindenberg, has been used as a reference for
radiosounding measurements to understand conventional RAOB sparseness
uncertainty. In fact this GRUAN station is also a conventional RAOB station
and, although the instrument is physically the same for both, data are
processed in a different manner giving different measurements. As a result for
temperature (WVMR) we have $K_{G}=306$ ($439$) GRUAN-RAOB co-locations.

Other meteorological variables considered in statistical modelling of the mismatch error, like wind, solar radiation or
geopotential, have been taken from the ERA-Interim global atmospheric
reanalysis implemented by the European Centre for Medium-Range Weather
Forecasts (ECMWF), see Berrisford et al. (2011).

\subsection{IASI}

Products retrieved from EUMETSAT's IASI instruments aboard of MetOp-A and
Met-Op B satellites considered in this study are based on version 6 of the
IASI level 2 processor. The infrared atmospheric sounding interferometer
(IASI) is a Fourier transform spectrometer based on the Michelson
interferometer, associated with an integrated imaging system (Blumstein et
al., 2004).

The IASI atmospheric profiles of temperature (WVMR) are available at 74
$\left(  32\right)  $ pressure levels in the range $958.6-10$ $hPa$
($958.6-300$ $hPa$). Considering vertical smoothing, the IASI sounding
products represent thermodynamic states of deep atmospheric layers at variable
depths, due to the integrating nature of the radiation measurements at the top
of the atmosphere. The maximum number of independent pieces of information is
approximately 14 (10) for temperature (humidity) profiles, the exact number
depending on atmospheric conditions. Hence the true vertical resolution is
quite lower than the vertical grid of 74 $\left(  32\right)  $ pressure levels
discussed above and profiles retrieved from such radiance measurements are
smoothed versions, where the smoothing functions are given by the so-called
averaging kernels. Although version 6 of the IASI level 2 processor provides
the information to calculate the averaging kernels, it has not been used in
this paper, being not available in the NPROVS data set.

\subsection{Conventional RAOB\label{sec:RAOB.data}}

Conventional radiosonde observations have been used historically as a de facto
standard data set in satellite calibration (via radiative transfer models) and
validation. Worldwide there are more than 2000 radiosonde launch sites and
mobile ship-based launch station.

While a radiosonde transmits an essentially continuous stream of temperature
and humidity information back to the station (each 5-10 m of altitude,
measured each 1-2 s), for temperature (WVMR), RAOB data set includes only 15
(6) "mandatory levels" in the atmospheric range $958.6-10$ $hPa$ ($958.6-300$
$hPa$). Moreover data are given at various "significant levels" which are the
pressure levels where a significant change or an extreme is identified in the
vertical temperature and/or dewpoint temperature profiles. For this reason,
the ECV variation between two such significant level is often assumed close to
linear. In practice, altitude and number of significant levels change among
different profiles and, on average, 28 significant levels per profile are
available in the RAOBs collected at NPROVS, the exact numbers depending on
specific atmospheric conditions.

\subsection{GRUAN}

Conventional RAOBs may not be able to provide reference-quality in situ and
ground-based remote sensing observations of upper-air essential climate
variables for metrological and traceability reasons, see Seidel et al.
$\left(  2009\right)  $, Immler et al. $\left(  2010\right)  $ and Bojinski et
al. $\left(  2014\right)  $. Improving on this, GRUAN data processing was
developed to meet the criteria for reference measurements (Dirksen et al.,
2014). As a result, GRUAN radiosounding profiles are provided together with
individual measurement uncertainty estimates at high vertical/temporal
resolution: measurements are obtained at 1-2 s or 5-10 m in altitude; this
temporal resolution is then reduced to about 10 s during
processing by a low pass filter to avoid temperature spikes. GRUAN quality has
been extensively assessed, see e.g. Calbet et al. $\left(  2017\right)  $.
Since Lindenberg GRUAN station is also a conventional RAOB station, it is
important to remark here that, considering this station, the two profiles
differ for the vertical resolution and for data processing. In fact the former
is obtained using the GRUAN processing algorithm while the latter is obtained
using the algorithm implemented in Vaisala RS92 instruments. As a result, the
two products give non coinciding measurements.

\section{Co-location mismatch sources\label{sec:mismatches}}

As above discussed, the comparison of radiosonde and IASI profiles aims at
understanding which factors contribute to the discrepancies observed between a
satellite vertical profile and a comparator profile. In principle the
comparator is taken as an error free "true" state, but, in practice, its
uncertainty is worth to be considered.

In fact a meaningful comparison should take into account: the spatio-temporal
mismatch between profiles; the different vertical smoothing and resolution of
the two instruments/data sets; the different horizontal resolution of the two
instruments. Since the latter is not important for IASI which has a relatively
small footprint, the following subsections briefly discuss the former points
and call for a comparison based on data harmonization.

\subsection{Satellite vertical smoothing}

As discussed in Section \ref{sec:datasets}, radiosonde and IASI\ are based on
completely different measurement techniques. While the radiosonde is able to
make a "direct" measurement of the ECV at the position in space and time
reached by the weather balloon, IASI sounds the atmosphere using an
interferometry technique. This implies that the vertical resolution of IASI is
much lower than the resolution characterizing a radiosonde. Any comparison
between radiosonde and IASI profiles, thus, may be affected by these differences.

Note that different methods are available in satellite product validation to
resolve the issue in vertical resolution difference. One requires to apply
satellite sounder averaging kernel to the target data, e.g., radiosonde data,
and then to compare with the retrievals (Maddy et al. 2008 ); one is to first average
vertical layer for both satellite and target data profiles, and
then to compute the validation statistics at those \textquotedblleft
coarse\textquotedblright\ vertical layers (Tobin et al. 2006; Sun et al.
2017). In this study, we choose to employ the vertical harmonization technique
for the uncertainty assessment, see Section \ref{sec:stat.harmonization}.

\subsection{Comparator uncertainty and vertical sparseness}

It has been seen that RAOB data are at sub-reference level and are provided at
the so-called mandatory and significant pressure levels, which are sparse
vertically. The latter being given at pressure levels where some interesting
variation is happening. This entails that data occurs according to a
preferential sampling design (Diggle et al., $2010$) which is dependent on
second order derivatives.

As a consequence, before developing a IASI comparison, the estimation of the
"true" profile at any pressure level, based on RAOB data, requires a
statistical assessment. In this frame, (vertical) sparseness uncertainty is
the uncertainty component related to the coarse vertical resolution.

\subsection{Spatio-temporal mismatch}

Radiosonde and IASI profiles are characterized by a spatio-temporal mismatch.
For the data set considered, only co-locations with horizontal distances up to
$300~$km at surface and time delays up to three hours have been considered,
see Kursinski and Hajj $\left(  2001\right)  $. This is because it is
impossible to perfectly synchronize the weather balloon launch with the
satellite overpass, and the overpass may be far from the station where the
radiosonde is launched, both in space and time. Additionally, the IASI profile
is retrieved nearly instantly while weather balloons take on average 1.7 hrs
from surface up to 10 hPa (Seidel at el. 2011). Moreover, the latter is
shifted by winds during its ascent. This means that profiles are never
perfectly co-located even if the satellite nadir viewing overpasses the launch station.

\section{Statistical harmonization and
uncertainties\label{sec:stat.harmonization}}

Vertical harmonization refers to a data transformation, which reduces the
differences in the vertical smoothing between the two profiles and improves the
radiosonde and IASI profiles comparability. In our case, the low vertical
resolution of IASI implies that IASI retrievals are much smoother than
radiosonde data. Since we cannot un-smooth IASI profiles, the radiosonde
profiles are smoothed in order to mimic the IASI retrievals.

This is given by the convolution of the radiosonde profile $s\left(  p\right)
$ of temperature and humidity with a normalized weighting function $w\left(
p;p^{\prime}\right)  $ in the pressure range $\left(  \dot{p},\ddot{p}\right)
$. Namely%

\begin{equation}
\tilde{s}\left(  p\right)  =\int_{\dot{p}}^{\ddot{p}}s\left(  q\right)
w\left(  q;p\right)  dq \label{eq:harmonization1}%
\end{equation}
with $\int_{\dot{p}}^{\ddot{p}}w\left(  q;p\right)  dq=1.$

The "true" profile may be assumed a continuous function of pressure, but RAOB
profiles are observed only at a limited number of "preferential" levels. To
handle this a two steps procedure is proposed. The first step, developed in
Section \ref{sec:RAOB.estimation}, extends the idea of Fass\`{o} et al.
$\left(  2014\right)  $ to represent atmospheric profiles as functional data
(Ramsay and Silverman, $2002$) with smoothness coefficient obtained by
minimizing the difference with the reference GRAUN data. The second step,
described in Sections \ref{sec:vertical.smoothing} and
\ref{sec:vertical.smoothing.uncertainty}, optimizes the weighting function $w$.

\subsection{Data and likelihood function}

Let us consider a collection of $K$ co-located RAOB-IASI profiles observed
across the geographic area and time frame of interest, with a subset of these
$K$ co-locations, say $\left\{  1,\ldots,K_{G}\right\}  ,$ obtained at
Lindenberg station, and having also GRUAN profile counterparts.

For each given co-location $k=1,\ldots,K$, let $\mathbf{x}_{J,k}$ be the
radiosonde data vector related to pressure levels $\mathbf{p}_{J,k}=\left(
p_{J,k,1},\ldots,p_{J,k,N_{J,k}}\right)  $, with $J=R$ for RAOB or $J=G$ for
GRUAN. Moreover let $\mathbf{x}_{I,k}$ be the IASI data vector related to
pressure levels $\mathbf{p}_{I}=\left(  p_{I,1},\ldots,p_{I,M}\right)  $. Note
that the RAOB pressure levels depends on co-location $k$. Instead IASI
pressure levels are invariant among co-locations with $M=74$ (32) for
temperature (WVMR). As a consequence RAOB and IASI pressure levels are
different and vertical matching may represent an issue. On the contrary, the
number of GRUAN measurements $N_{G,k}$ is very high for all profiles, so that
for any prefixed RAOB level in $\mathbf{p}_{R,k},$ a very close GRUAN pressure
level in $\mathbf{p}_{G,k}$ may be found.

\subsection{RAOB estimation and sparseness
uncertainty\label{sec:RAOB.estimation}}

The discrepancies between conventional RAOB and GRUAN may be used to
understand the loss of information of RAOB due to its sparse vertical
resolution. In fact, the minimization of this loss can be used to define an
optimal estimate of the unobserved true signal.

To see this, the true signal of $k-th$ profile is considered as a smooth
function denoted by $s_{k}^{0}\left(  p\right)  $, and it is related to
observation $x_{J,k}\left(  p\right)  $ for $p\in\mathbf{p}_{J}$ and $J=R,G,I$
by the following conditions
\begin{equation}
x_{J,k}\left(  p\right)  =s_{J,k}\left(  p\right)  +\varepsilon_{J,k}\left(
p\right)  \label{eq:measurement.error}%
\end{equation}
where $\varepsilon_{J,k}\left(  p\right)  $ is Gaussian distributed, $N\left(
0,\sigma_{J,k}^{2}\left(  p\right)  \right)  $, and
\begin{align}
s_{G,k}\left(  p\right)   &  =s_{k}^{0}\left(  p\right) \nonumber\\
s_{R,k}\left(  p\right)   &  =s_{k}^{0}\left(  p\right)  +\Delta\left(
p\right) \label{eq:RAOB.bias}\\
s_{I}\left(  p\right)   &  =\int s_{k}^{0}\left(  q\right)  w\left(
q;p\right)  dq. \label{eq:IASI.integral}%
\end{align}

These three conditions will be discussed in details later. For the moment note
that $\Delta\left(  p\right)  $ is a smooth bias, constant over co-locations
and $w$ is a weighting function. Moreover, note that, for GRUAN, the squared
measurement uncertainty $u_{G}^{2}\left(  p\right)  =E\left(  x_{G}\left(
p\right)  -s^{0}\left(  p\right)  \right)  ^{2}$ is known at all pressure
levels $p\in\mathbf{p}_{G}$ and $u_{G}^{2}=\sigma_{G}^{2}$. For RAOB the
measurement uncertainty $u_{R}^{2}=\sigma_{R}^{2}+\Delta^{2}$ is not widely
available but there is some evidence that $\sigma_{R,j}^{2}\left(  p\right)
\cong\rho\sigma_{G,j}^{2}\left(  p\right)  $ for some $\rho\geq1$. For
simplicity we assume tha $\rho$ does not depend on pressure level $p$ nor on
co-location $k$.

In this paper, we estimate the smooth profile $s_{R}\left(  p\right)  $ by
$\hat{s}_{R}\left(  p,\lambda\right)  $ which is a penalized spline with
smoothing factor $\lambda$. The estimated profile $\hat{s}_{R}\left(
p;\lambda\right)  $ is computed on RAOB data by solving the following
penalized weighted least square problem:%
\begin{align}
\hat{s}_{R,k}\left(  p;\lambda\right)   &  =\arg\min_{s}\left[  \sum
_{j=1}^{N_{R,k}}\right.  \left(  x_{R,k}\left(  p_{j}\right)  -s\left(
p_{j}\right)  \right)  ^{2}\alpha_{R,k}\left(  p\right)
\label{eq:RAOB.WLS.criterion}\\
&  +\lambda\left.  \sum_{j=1}^{N_{R,k}}\left(  \frac{\partial^{2}}{\partial
p^{2}}s\left(  p_{j}\right)  \right)  ^{2}\alpha_{R,k}\left(  p\right)
\right] \nonumber
\end{align}
where
\begin{equation}
\alpha_{R,k}\left(  p\right)  =\alpha_{G,k}\left(  p\right)  =u_{G,k}\left(
p\right)  ^{-2}/\sum_{k=1}^{K_{G}}u_{G,k}\left(  p\right)  ^{-2}
\label{eq:uncertainty_weight_GRUAN}%
\end{equation}
for co-locations in Lindeberg and $\alpha_{R}=\frac{1}{N_{R,k}}$ else.

Following e.g. Reinsch $\left(  1971\right)  $, and using one knot per
observation, the solution $\hat{s}_{R,k}$ of equation $\left(
\ref{eq:RAOB.WLS.criterion}\right)  $ may be expressed in terms of tolerance
$\tau=\tau\left(  \lambda\right)  $, which is the upper limit of weighted root
mean squared error along the RAOB profile
\begin{equation}
\frac{1}{N_{R,k}}\sum_{p\in\mathbf{p}_{R,k}}\left\vert x_{R,k}\left(
p\right)  -\hat{s}_{R}\left(  p,\lambda\right)  \right\vert ^{2}\alpha
_{R,k}\left(  p\right)  \leq\tau^{2} \label{eq:tau2}%
\end{equation}
where, clearly, $\tau=0$ gives interpolating splines. For this reason,
depending on the context, we will use either $\tau$ or $\lambda$ to address
smoothing properties of spline $\hat{s}_{R,k}\left(  p,\lambda\right)
=\hat{s}_{R,k}\left(  p,\tau\right)  $.

Using GRUAN-RAOB comparison at Lindenberg, we estimate the bias $\Delta$ by
the weighted RAOB-GRUAN average difference, namely%
\begin{equation}
\hat{\Delta}\left(  p,\tau\right)  =\sum_{k=1}^{K_{G}}\left(  \hat{s}%
_{R}\left(  p;\tau\right)  -x_{G,k}\left(  p\right)  \right)  \alpha
_{G,k}\left(  p\right)  . \label{eq:Delta.hat}%
\end{equation}
Next, the smoothing factor $\lambda\left(  \tau\right)  $ is obtained by
optimizing the adjusted GRUAN-RAOB difference. In other words, $\tau$ is the
solution of the following weighted least squares criterion:%
\begin{equation}
\hat{\tau}=\arg\min_{\tau}\sum_{k=1}^{K_{G}}\sum_{p\in\mathbf{P}_{G,k}}\left[
x_{G,k}\left(  p\right)  -\left(  \hat{s}_{R}\left(  p;\tau\right)
-\hat{\Delta}\left(  p;\tau\right)  \right)  \right]  ^{2}\alpha_{G,k}\left(
p\right)  \label{eq:gruan.WLS.criterion}%
\end{equation}
where $\alpha_{G}$ is defined in Equation $\left(
\ref{eq:uncertainty_weight_GRUAN}\right)  $.

Due to the peculiarity of the RAOB sampling points discussed in Section
\ref{sec:RAOB.data}, three spline models are compared in Section
\ref{sec:Case.Study}: linear and cubic smoothing Bsplines and Hermite
interpolating splines (Hsplines). The former two are well known and we only
remark here that the smoothing coefficient $\tau$ is not obtained by a
cross-validation $\left(  \text{CV}\right)  $ or generalized CV criterion on
RAOB data as in standard smoothing splines. Instead $\tau$ is numerically
optimized according to the GRUAN agreement criterion $\left(
\ref{eq:gruan.WLS.criterion}\right)  $ which takes into account measurement
uncertainty. Interpolating Hsplines are also known as piecewise cubic Hermite
interpolating polynomials, being cubic monotonic splines with continuous first
derivatives, see Fritsch and Carlson $\left(  1980\right)  .$\ Hence Hsplines
are introduced here as an intermediate solution between cubic and linear
Bsplines. In fact, in this model selection problem, one could use also
smoothing Hsplines, this approach being further discussed in the case study.

After obtaining $\hat{\tau}$, the optimized quantity in Equation $\left(
\ref{eq:gruan.WLS.criterion}\right)  $ provides the total mismatch uncertainty
profile of RAOB-GRUAN comparison, namely:%
\begin{equation}
u_{RG.tot}^{2}\left(  p\right)  =\sum_{k=1}^{K_{G}}\left[  x_{G,k}\left(
p\right)  -\hat{s}_{R}\left(  p;\hat{\tau}\right)  \right]  ^{2}\alpha
_{G,k}\left(  p\right)  \label{eq:u_RG_total}%
\end{equation}
which is the loss of information due to sparseness and difference in data
processing. More comments and the decomposition of Equation
\ref{eq:u_RG_total} in sparseness and processing uncertainty will be developd
in the case study.

\subsection{Vertical smoothing\label{sec:vertical.smoothing}}

With the aim of making radiosonde and IASI profiles comparable, the radiosonde
profile of the previous section, denoted in the sequel by $\hat{s}_{R}\left(
p\right)  =\hat{s}_{R}\left(  p;\hat{\tau}\right)  $ is smoothed by means of a
weighted integral, namely:%
\begin{equation}
\tilde{s}_{R}\left(  p;\boldsymbol{\theta}\right)  =\int_{p_{\operatorname*{L}%
}}^{\ddot{p}}\hat{s}_{R}\left(  q\right)  w\left(
q;\mathbf{\boldsymbol{\theta}},p\right)  dq. \label{eq:smoothing}%
\end{equation}
In Equation $\left(  \ref{eq:smoothing}\right)  $, the weight function
$w\left(  \cdot;\mathbf{\boldsymbol{\theta}},p\right)  $ is a non negative and
normalized weight function, depending on $p$ and a parameter vector
\textbf{$\boldsymbol{\theta}$}$=$\textbf{$\boldsymbol{\theta}$}$\left(
p\right)  $, which has to be estimated. Since the role of $w$ is to mimic the
IASI sounding of the atmosphere, Fass\`{o} et al. $\left(  2017\right)  $
considered the following alternative functions: rectangular, sine, Gaussian
and GEV distribution. As expected the latter was found to outperform the other
simpler competitors and, for this reason, it is used here. In particular the
GEV pdf has parameter vector $\left(  \mu,\sigma,\xi\right)  $, which are the
location, scale and shape parameter, respectively, see Kotz and Nadarajah
$\left(  2000\right)  .$ In this paper we use level dependent parameters,
namely $\mu\left(  p\right)  =p$ and \textbf{$\boldsymbol{\theta}$}$\left(
p\right)  =\left(  \sigma\left(  p\right)  ,\xi\left(  p\right)  \right)  \,$.

In order to compute the harmonized RAOB $\tilde{s}$ from Equation
\ref{eq:smoothing}, we need to estimate \textbf{$\boldsymbol{\theta}$} and a
natural choice is the following penalized weighted least squares iterated for
$j=1,\ldots,M:$
\begin{align}
\mathbf{\boldsymbol{\hat{\theta}}}_{j}  &  \mathbf{=\boldsymbol{\hat{\theta}}%
}\left(  p_{j}\right)  =\arg\min_{\boldsymbol{\theta}}\left[
%TCIMACRO{\dsum \limits_{k=1}^{K}}%
%BeginExpansion
{\displaystyle\sum\limits_{k=1}^{K}}
%EndExpansion
\right.  \left[  x_{I,k}\left(  p_{j}\right)  -\tilde{s}_{R,k}\left(
p_{j};\mathbf{\boldsymbol{\theta}}\right)  \right]  ^{2}\alpha_{I,k}%
^{2}\left(  p_{j}\right) \label{eq:GEV.minimization}\\
&  +\left.  I\left(  j>1\right)  \left\Vert \mathbf{\boldsymbol{\theta}%
}-\mathbf{\boldsymbol{\hat{\theta}}}_{j-1}\right\Vert _{\Sigma_{\zeta}%
}\right]  \text{,}\nonumber
\end{align}
where $p_{j}\in\mathbf{p}_{I}$ and the weight $\alpha_{I}^{2}$ is the
normalized squared reciprocal measurement uncertainty of IASI, analogous to
formula $\left(  \ref{eq:uncertainty_weight_GRUAN}\right)  $. Moreover
$\left\Vert x\right\Vert _{\Sigma}=x^{\prime}\Sigma^{-1}x\,,$where
$\Sigma_{\zeta}$ is a variance covariance matrix to be discussed in the next
section and $I\left(  j>1\right)  =1$ if $j>1$ and $=0$ else. Note that the
penalty term in (\ref{eq:GEV.minimization}) is related to smoothness of the
atmosphere and hence of \textbf{$\boldsymbol{\theta}$}$\left(  p\right)  $ wrt
$p$.

\subsection{Vertical smoothing
uncertainty\label{sec:vertical.smoothing.uncertainty}}

A byproduct of the data harmonization procedure above described is the
uncertainty component related to the vertical smoothing. In particular, the
RAOB-IASI mismatch uncertainty due to difference in vertical smoothing is
given by%
\begin{equation}
u_{RI.vsmooth}^{2}\left(  p\right)  =u_{RI.raw}^{2}\left(  p\right)
-u_{RI.harm}^{2}\left(  p\right)  \label{eq:VS.u.decomposition}%
\end{equation}
where $u_{RI.harm}\left(  p\right)  $ is the vertically harmonized mismatch
uncertainty:%
\begin{equation}
u_{RI.harm}^{2}\left(  p\right)  =%
%TCIMACRO{\dsum \limits_{k=1}^{K}}%
%BeginExpansion
{\displaystyle\sum\limits_{k=1}^{K}}
%EndExpansion
\left[  \tilde{s}_{k}\left(  p;\mathbf{\boldsymbol{\hat{\theta}}}\left(
p\right)  \right)  -x_{I,k}\left(  p\right)  \right]  ^{2}\alpha_{I,k}\left(
p\right)  \label{eq:u.harmonized}%
\end{equation}
and $u_{RI.raw}\left(  p\right)  $ is the raw mismatch uncertainty in the
comparison of IASI with non harmonized RAOB $s_{R}\left(  p\right)  ,$ namely%
\[
u_{RI.raw}^{2}\left(  p\right)  =\frac{1}{K}%
%TCIMACRO{\dsum \limits_{k=1}^{K}}%
%BeginExpansion
{\displaystyle\sum\limits_{k=1}^{K}}
%EndExpansion
\left[  x_{I,k}\left(  p\right)  -\hat{s}_{R,k}\left(  p\right)  \right]
^{2}\alpha_{I,k}\left(  p\right)  .
\]
Hence, $u_{RI.vsmooth}^{2}\left(  p_{j}\right)  $ may be interpreted as the
(squared) mismatch uncertainty improved by the data harmonization.

\section{Likelihood inference\label{sec:likelihood}}

The modeling machinery of the previous section has a rigorous interpretation
as a maximum likelihood estimation problem for a non linear mixed effect
model. This is properly described using three main steps for RAOB, GRUAN and
IASI respectively.

\subsection{RAOB likelihood}

The first step is to represent the RAOB true signal by the following linear
representation%
\begin{equation}
s_{R}\left(  p\right)  =\mathbf{B}\left(  p\right)  ^{\prime}%
\mathbf{\boldsymbol{\gamma}}_{R} \label{eq:Bspline}%
\end{equation}
where $\mathbf{B}$ is the vector of Bspline basis functions and
\textbf{$\boldsymbol{\gamma}$}$_{R}$ is the vector of the spline coefficients.
Using this, Equation $\left(  \ref{eq:measurement.error}\right)  $ for $k-th$
RAOB profile may be rewritten as follows:%
\[
x_{R,k}\left(  p\right)  =\mathbf{B}\left(  p\right)  ^{\prime}%
\mathbf{\boldsymbol{\gamma}}_{R,k}+\varepsilon_{R,k}\left(  p\right)
\]
where $p\in\mathbf{p}_{R,k},k=1,\ldots,K$.

Stacking $x_{R,k}\left(  p_{1}\right)  ,\ldots,x_{R,k}\left(  p_{N_{R,k}%
}\right)  $ in a vector, say $X_{R,k}$, all $\mathbf{B}^{\prime}s$ in a matrix
$Z_{R,k}$ and all errors in a vector $\mathbf{\varepsilon}$ we have the
following matrix representation%
\[
X_{R,k}=Z_{R,k}\boldsymbol{\gamma}_{R,k}+\mathbf{\varepsilon}_{R,k}.
\]
This has a well known mixed effects model interpretation (see e.g. Fahrmeir et
al., $2013$). To see this, let $\left[  \boldsymbol{\gamma}\right]  $ denotes
the probability distribution of the random vector $\boldsymbol{\gamma},$ and
assume that, $\left[  \boldsymbol{\gamma}_{R,k}\right]  =N\left(
0,G_{k}\right)  $ where $G_{k}=\lambda^{-2}I_{k}$, $\lambda>0$ is a known
smoothing factor and $I_{K}$ is the identity $k-\dim$ matrix. Than the
penalized least square criterion in Equation $\left(
\ref{eq:RAOB.WLS.criterion}\right)  $ corresponds to the maximum likelihood
estimate (MLE). In fact, apart for an additive constant, we have%

\[
-2\log\left[  \boldsymbol{\gamma}_{k}\right]  \left[  X_{R,k}%
|\boldsymbol{\gamma}_{k}\right]  =\left\Vert \boldsymbol{\gamma}%
_{k}\right\Vert _{G_{k}}+\left\Vert X_{R,k}-Z_{R,k}\boldsymbol{\gamma}%
_{k}\right\Vert _{\Sigma_{R}}.
\]
Stacking all $X_{R,k}$ in a vector $X_{R}$, and similarly for
$Z,\boldsymbol{\gamma}$ and $\varepsilon$, we have the following matrix
representation%
\[
X_{R}=Z_{R}\mathbf{\boldsymbol{\gamma}}+\mathbf{\varepsilon}_{R}%
\]
and%
\begin{align*}
-2\log\left[  \mathbf{\boldsymbol{\gamma}}\right]  \left[  X_{R}%
|\boldsymbol{\gamma}\right]   &  =\left\Vert \mathbf{\boldsymbol{\gamma}%
}\right\Vert _{G}+\left\Vert X_{R}-Z_{R}\mathbf{\boldsymbol{\gamma}%
}\right\Vert _{\Sigma_{R}}\\
&  =\sum_{k=1}^{K}\left(  \left\Vert \boldsymbol{\gamma}_{k}\right\Vert
_{G_{k}}+\left\Vert X_{R,k}-Z_{R,k}\boldsymbol{\gamma}_{k}\right\Vert
_{\Sigma_{R}}\right)
\end{align*}
where $\Sigma_{R}$ is diagonal matrix corresponding to uncertainties in
Equation \ref{eq:uncertainty_weight_GRUAN}. Since this likelihood is optimized
by minimizing each summand independently, the computation burden is linear in
$K$ and the solution is the MLE \textbf{$\boldsymbol{\hat{\gamma}}$}$\left(
\tau\right)  $ as a function of $\tau$ (or $\lambda$).

Hence Equation $\left(  \ref{eq:RAOB.WLS.criterion}\right)  $ may be rewritten as%

\[
\hat{s}_{R,k}\left(  p;\tau\right)  =\mathbf{B}\left(  p\right)  ^{\prime
}\mathbf{\boldsymbol{\hat{\gamma}}}_{R}\left(  \tau\right)  \text{.}%
\]

Notice that the RAOB data model for $X_{R}$ may be partitioned as
\[
X_{R}=\left(
\begin{array}
[c]{c}%
X_{R_{1}}\\
X_{R_{2}}%
\end{array}
\right)  =\left(  Z_{R_{1}},Z_{R_{2}}\right)  \left(
\begin{array}
[c]{c}%
\mathbf{\boldsymbol{\gamma}}_{R_{1}}\\
\mathbf{\boldsymbol{\gamma}}_{R_{2}}%
\end{array}
\right)  +\left(
\begin{array}
[c]{c}%
\mathbf{\varepsilon}_{R_{1}}\\
\mathbf{\varepsilon}_{R_{2}}%
\end{array}
\right)
\]
where $R_{1}$ is the GRUAN matching data set corresponding to co-locations
$k=1,\ldots,K_{G}$, and $R_{2}\ $is the remaining major part of the RAOB data
set with $K-K_{G}$ soundings.

\subsection{GRUAN-RAOB likelihood}

Now, considering GRUAN and RAOB matching data in $R_{1}$, we have the
following representation%
\begin{align*}
x_{G}\left(  p\right)   &  =s_{G}\left(  p\right)  +\varepsilon_{G}\left(
p\right) \\
s_{G}\left(  p\right)   &  =s_{R}\left(  p\right)  +\Delta\left(  p\right)  ,
\end{align*}
where $\Delta$ is a GRUAN-RAOB fixed effect bias. Hence, in matrix notation,
we may write%

\[
X_{G}=Z_{R_{1}}\boldsymbol{\gamma}+\Delta+\varepsilon_{G}%
\]
and%
\[
-2\log\left[  X_{G}|X_{R_{1}},\boldsymbol{\gamma}_{R_{1}}\right]  =\left\Vert
X_{G}-Z_{G}\boldsymbol{\gamma}_{R_{1}}-\Delta\right\Vert _{\Sigma_{G}}.
\]

If we compute this at $\boldsymbol{\hat{\gamma}}$$\left(  \tau\right)  $ we
have a profile log-likelihood $l\left(  \tau,\Delta|\boldsymbol{\hat{\gamma}%
}\left(  \tau\right)  \right)  $ which is easily optimized for $\Delta\left(
\tau\right)  $ and finally for $\left(  \tau|\hat{\Delta}\left(  \tau\right)
,\boldsymbol{\hat{\gamma}}\left(  \tau\right)  \right)  .$ We than have the
MLE triplet for IASI and RAOB data sets:%
\[
\left(  \boldsymbol{\hat{\gamma}},\hat{\Delta},\hat{\tau}\right)
\]
which is given by Equations $\left(  \ref{eq:RAOB.WLS.criterion}\right)
,\left(  \ref{eq:Delta.hat}\right)  $ and $\left(
\ref{eq:gruan.WLS.criterion}\right)  $.

\subsection{IASI-RAOB likelihood\label{sec:IASI.RAOB.likelihood}}

The IASI observation equation is obtained by substituting Equations $\left(
\ref{eq:RAOB.bias}\right)  ,\left(  \ref{eq:Bspline}\right)  $ and $\left(
\ref{eq:IASI.integral}\right)  $ in Equation $\left(
\ref{eq:measurement.error}\right)  .$ This gives%
\begin{equation}
x_{I,k}\left(  p_{j}\right)  =Z_{I,k}\left(  p_{j},\mathbf{\boldsymbol{\theta
}}_{j}\right)  \boldsymbol{\gamma}_{k}-\tilde{\Delta}\left(  p_{j}%
,\mathbf{\boldsymbol{\theta}}_{j}\right)  +\varepsilon_{I}\text{.}
\label{eq:smoothing_2}%
\end{equation}
where $Z_{I,k}\left(  p,\mathbf{\boldsymbol{\theta}}\right)  =\int_{\dot{p}%
}^{\ddot{p}}w\left(  q;\mathbf{\boldsymbol{\theta}},p\right)  B_{R,k}\left(
q\right)  ^{\prime}dq$ and $\tilde{\Delta}\left(  p,\mathbf{\boldsymbol{\theta
}}\right)  =\int_{\dot{p}}^{\ddot{p}}w\left(  q;\mathbf{\boldsymbol{\theta}%
},p\right)  \Delta\left(  q\right)  dq$.

Hence the stacked IASI observation equation for $j-th$ pressure level and all
co-locations may be written as follows:%
\[
X_{I.j}=Z_{I,j}\left(  \mathbf{\boldsymbol{\theta}}_{j}\right)
\mathbf{\boldsymbol{\gamma}}-\mathbf{\tilde{\Delta}}\left(
\mathbf{\boldsymbol{\theta}}_{j}\right)  +\mathbf{\varepsilon}_{I,j}%
\]
and the full data set is represented by $X_{I}=Z_{I}\left(
\mathbf{\boldsymbol{\Theta}}\right)  $\textbf{$\boldsymbol{\gamma}$%
}$-\mathbf{\tilde{\Delta}}\left(  \boldsymbol{\Theta}\right)
+\mathbf{\varepsilon}_{I},$ where $\boldsymbol{\Theta=}\left(
\mathbf{\boldsymbol{\theta}}_{1},\ldots,\mathbf{\boldsymbol{\theta}}%
_{M}\right)  $.

Now, in order to estimate $\boldsymbol{\Theta}$, one could consider
independent estimates for \textbf{$\boldsymbol{\theta}$}$_{j}\ $separately.
But this assumption contrasts with atmospheric considerations and tends to
overfit. On the opposite side, one could assume
\textbf{$\boldsymbol{\theta}$}$\left(  p\right)  $ is a smooth function of $p$
and use Bspline. This would largely increase the number of parameters to be
simultaneously optimized, resulting in an unfeasible algorithm. An
intermediate and suitable solution is to assume that
\textbf{$\boldsymbol{\theta}$} is a vector random walk, namely%
\begin{equation}
\mathbf{\boldsymbol{\theta}}_{j}=\mathbf{\boldsymbol{\theta}}_{j-1}+\zeta_{j}
\label{eq:random.walk}%
\end{equation}
for $j=2,\ldots,M.$ In the equation above, \textbf{$\boldsymbol{\theta}$}%
$_{1}$ is an unknown parameter and the innovations $\zeta_{j}$ are Gaussian
distributed $N\left(  0,\Sigma_{\zeta}\right)  $ with $\Sigma_{\zeta}$ a
diagonal matrix.

It follows that, the profile likelihood for $\boldsymbol{\gamma}%
=\boldsymbol{\hat{\gamma}}$, $\Delta=\hat{\Delta}$ and known $\Sigma_{\zeta}$
and $\Sigma_{I}$ is given by%
\begin{equation}%
\begin{tabular}
[c]{c}%
$-2\log\left[  \mathbf{\zeta},X_{I}|\boldsymbol{\gamma},X_{R,}X_{G}\right]
=\left\Vert Z_{I,1}\left(  \mathbf{\boldsymbol{\theta}}_{1}\right)
\mathbf{\boldsymbol{\gamma}}-\mathbf{\tilde{\Delta}}\left(
\mathbf{\boldsymbol{\theta}}_{1}\right)  \right\Vert _{\Sigma_{I}}+$\\
$+\sum_{j=2}^{M}\left(  \left\Vert \mathbf{\boldsymbol{\theta}}_{j}%
-\mathbf{\boldsymbol{\theta}}_{j-1}\right\Vert _{\Sigma_{\zeta}}+\left\Vert
Z_{I,j}\left(  \mathbf{\boldsymbol{\theta}}_{j}\right)
\mathbf{\boldsymbol{\gamma}}-\mathbf{\tilde{\Delta}}\left(
\mathbf{\boldsymbol{\theta}}_{j}\right)  \right\Vert _{\Sigma_{I}}\right)  $%
\end{tabular}
\ \ \ \ \ \ \label{eq:loglik.IASI}%
\end{equation}
which is optimized by the following $M$ nonlinear optimizations%

\begin{align}
\mathbf{\boldsymbol{\hat{\theta}}}_{1}  &  =\arg\min
_{\mathbf{\boldsymbol{\theta}}}\left\Vert Z_{I,1}\left(
\mathbf{\boldsymbol{\theta}}\right)  \mathbf{\boldsymbol{\gamma}%
}-\mathbf{\tilde{\Delta}}\left(  \mathbf{\boldsymbol{\theta}}_{1}\right)
\right\Vert _{\Sigma_{I}}\label{eq:GEV.minimization.MLE}\\
\mathbf{\boldsymbol{\hat{\theta}}}_{j}  &  =\mathbf{\boldsymbol{\hat{\theta}}%
}_{j-1}+\arg\min_{\mathbf{\zeta}}\left(  \left\Vert \mathbf{\zeta}\right\Vert
_{\Sigma_{\zeta}}+\left\Vert Z_{I,j}\left(  \mathbf{\boldsymbol{\hat{\theta}}%
}_{j-1}+\zeta\right)  \mathbf{\boldsymbol{\gamma}}-\mathbf{\tilde{\Delta}%
}\left(  \mathbf{\boldsymbol{\hat{\theta}}}_{j}\right)  \right\Vert
_{\Sigma_{I}}\right) \nonumber
\end{align}
for $j=2,\ldots,M$, which correspond to minimizations in Equation $\left(
\ref{eq:GEV.minimization}\right)  $. Note that the diagonal matrix $\Sigma
_{I}$ is given by the uncertainties in Equation $\left(
\ref{eq:GEV.minimization}\right)  $ which are assumed to be known up to an
acceptable approximation.

\section{Case study\label{sec:Case.Study}}

In this section, the two step harmonization procedure presented in Section
\ref{sec:stat.harmonization} is applied to the RAOB-IASI data set introduced
in Section \ref{sec:datasets}, independently for temperature and WVMR co-locations.

\subsection{RAOB estimation\label{sec:Case.Study:RAOB.estimation}}

The first step is the transformation of the sparse RAOB radiosonde profiles
into continuous functions to be used in the convolution of step two. To do
this, spline type and smoothing level have been chosen according to the GRUAN
closeness criterion of Equation $\left(  \ref{eq:gruan.WLS.criterion}\right)
$. Considering linear Bsplines, smoothing optimization is shown in Figures
\ref{fig:wrmse_by_tol_temp} and $\ref{fig:wrmse_by_tol_WVMR},$ giving
$\hat{\tau}=0.4K$ and $\hat{\tau}=0.075g/kg$ for temperature and humidity
respectively. Table \ref{tab:spline.selection} shows that linear Bsplines with
GRUAN-optimal smoothing improves over both cubic Bsplines and interpolating
Hsplines. This is consistent with the preferential sampling design
characterizing RAOB significant levels mentioned in Section
\ref{sec:RAOB.estimation}.

\begin{center}
\begin{table}[ptb]
\centering%
\begin{tabular}
[c]{lll}
& Temperature & WVMR\\\hline
Linear Bsplines & 0.4897 & 0.0346\\
Cubic Bsplines & 0.7706 & 0.0712\\
Interpolating Hsplines & 0.6085 & 0.0379\\\hline
\end{tabular}
\caption{Comparison of linear and cubic Bsplines and interpolating Hermite
splines (Hsplines) of temperature (K) and humidity (g/kg), based on weighted
root mean square error of RAOB-GRUAN in Lindenberg.}%
\label{tab:spline.selection}%
\end{table}

\begin{figure}[t]
\centering\includegraphics[width=10cm]{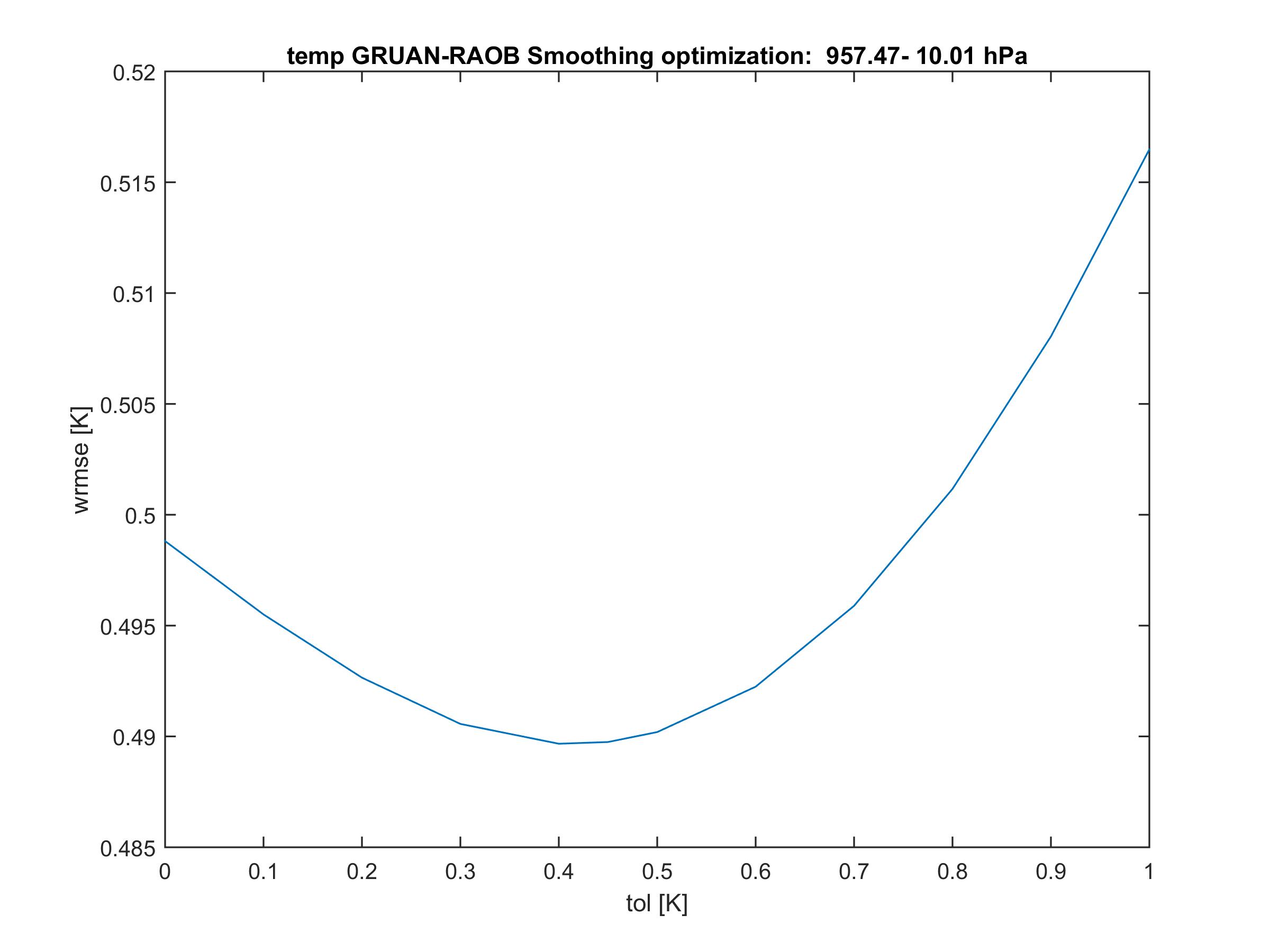}\caption{Temperature.
Smoothing optimization of RAOB linear Bsplines wrt GRUAN data in Lindenberg.
Abscissa: tolerance $\tau,$ given by Equation $\left(  \ref{eq:tau2}\right)
$. Ordinate: RAOB-GRUAN weighted root mean square error.}%
\label{fig:wrmse_by_tol_temp}%
\end{figure}

\begin{figure}[t]
\centering\includegraphics[width=10cm]{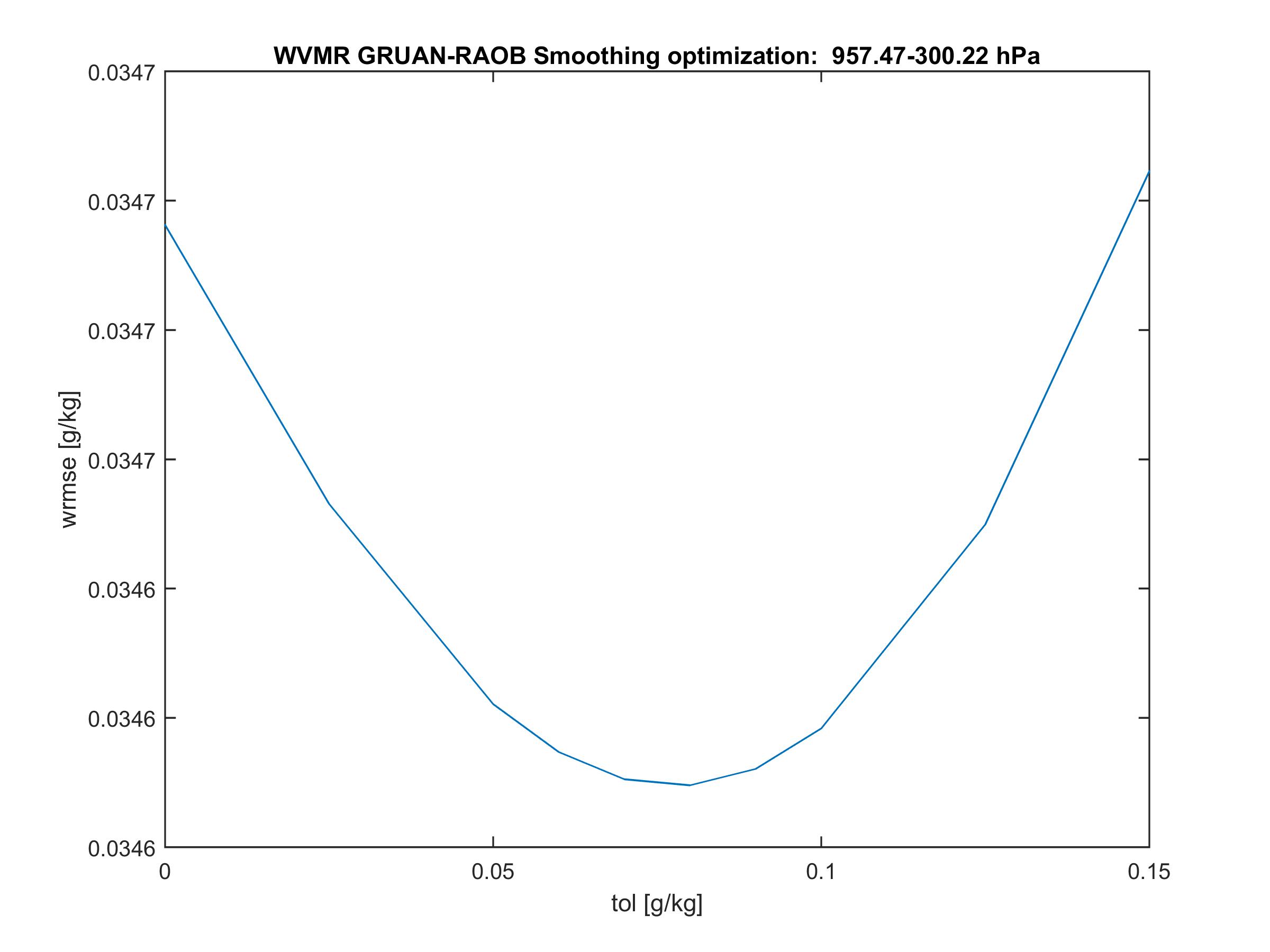}\caption{WVMR.
Smoothing optimization of RAOB linear Bsplines wrt GRUAN data in Lindenberg.
Abscissa: tolerance $\tau,$ given by Equation $\left(  \ref{eq:tau2}\right)
$. Ordinate: RAOB-GRUAN weighted root mean square error.}%
\label{fig:wrmse_by_tol_WVMR}%
\end{figure}\qquad\qquad\qquad
\end{center}

\subsection{Sparseness and processing uncertainty\label{sec:u_sparseness}}

\begin{figure}[t]
\centering\includegraphics[width=10cm]{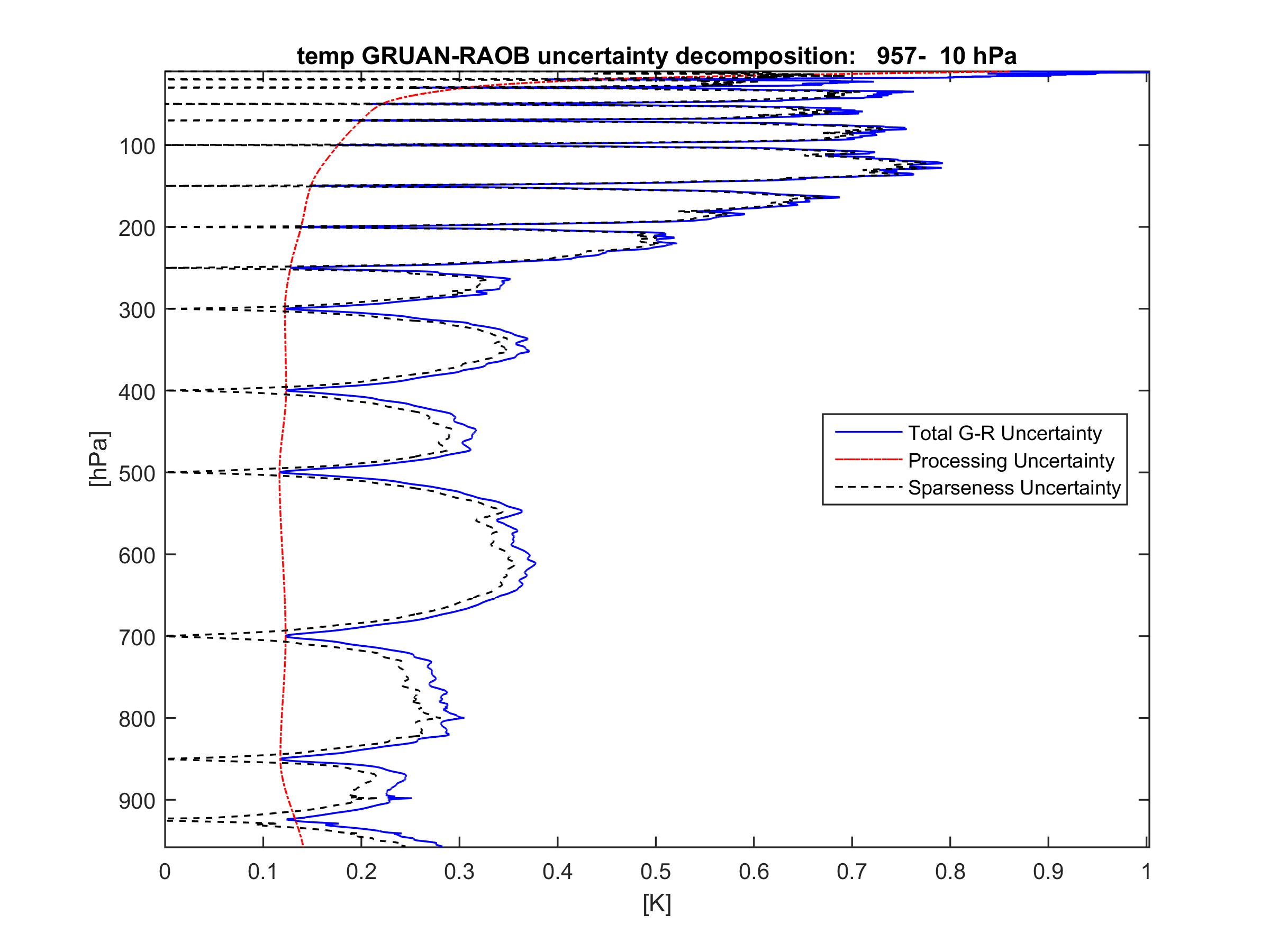}\caption{Temperature.
RAOB-GRUAN mismatch uncertainties: solid blu line is total mismatch
uncertainty $\left(  u_{RG.tot}\right)  $; solid red line is uncertainty due
to difference between Vaisala and GRUAN processing $\left(  u_{RG.proc}%
\right)  $; dashed black line is the RAOB sparseness uncertainty $\left(
u_{R.sparse}=\sqrt{u_{RG.tot}^{2}-u_{RG.proc}^{2}}\right)  $.}%
\label{fig:u_sparseness_temp}%
\end{figure}

\begin{figure}[t]
\centering\includegraphics[width=10cm]{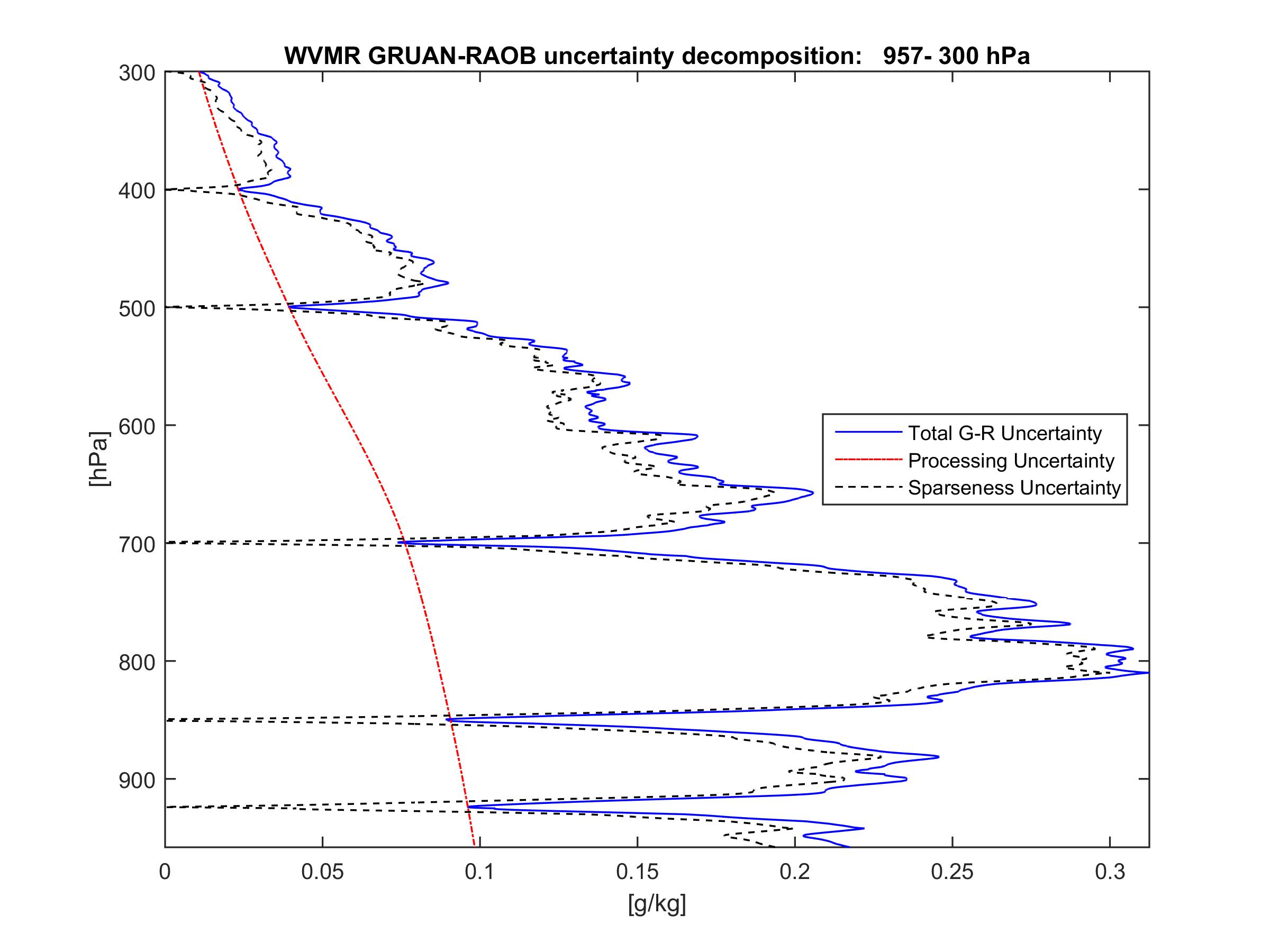}\caption{WVMR.
RAOB-GRUAN mismatch uncertainties: solid blu line is total mismatch
uncertainty $\left(  u_{RG.tot}\right)  $; solid red line is uncertainty due
to difference between Vaisala and GRUAN processing $\left(  u_{RG.proc}%
\right)  $; dashed black line is the RAOB sparseness uncertainty $\left(
u_{R.sparse}=\sqrt{u_{RG.tot}^{2}-u_{RG.proc}^{2}}\right)  $.}%
\label{fig:u_sparseness_wvmr}%
\end{figure}

The comparison of RAOB and GRUAN data in Lindenberg station provides the
GRUAN-RAOB total mismatch uncertainty, which is computed using the approach of
Section \ref{sec:RAOB.estimation}. In particular in Figures
\ref{fig:u_sparseness_temp} and \ref{fig:u_sparseness_wvmr}, $u_{RG.tot}$ of
Equation $\left(  \ref{eq:u_RG_total}\right)  $ shows a peculiar behavior with
local minima at mandatory levels. In fact, as discussed in Section
\ref{sec:datasets}, both RAOB and GRUAN are observed at these pressure levels,
while between them, RAOB is observed only at significant levels. The red line
interpolates between above minima and defines the mismatch due to difference
between GRUAN data processing (Dirksen et al, 2014) and Vaisala RS92 data
processing, denoted by $u_{RG.proc}$. As a result the black dashed line of
Figures \ref{fig:u_sparseness_temp} and \ref{fig:u_sparseness_wvmr} is the
sparseness uncertainty adjusted for mismatch in processing and is given by the
quadratic difference among the previous uncertainties, namely%
\[
u_{R.sparse}^{2}=u_{RG.tot}^{2}-u_{RG.proc}^{2}.
\]

Considering temperature the processing uncertainty is close to $0.1$ $K$ until
$300$ $hPa$. In this range also sparseness uncertainty is generally smaller
than $0.35$ $K$. In the upper atmosphere both uncertainties are larger
consistently with solar radiation bias. Considering WVMR, as expected, the
vertical pattern is reversed with a processing uncertainty decreasing from
$0.1$ $g/kg$ at ground level to $0.02$ $g/kg$ at $300$ $hPa.$ In this range
the sparseness uncertainty is smaller $0.3$ $g/kg.$

\subsection{Harmonization and vertical
smoothing\label{sec:vert.smoothing.estimation}}

Conventional RAOB profiles are harmonized thanks to the optimization in
Equation $\left(  \ref{eq:GEV.minimization.MLE}\right)  .$ This is solved for
each IASI pressure level $p_{j}\in\mathbf{p}_{I}$ iteratively from the top
pressure level $p_{1}=11$ $hPa$ $\left(  300\text{ }hPa\right)  $ and going
down to $p_{M}=957$ $hPa$ separately for temperature and WVMR. We tried also
to iterate in the opposite order, from ground to upper air, obtaining very
close results. At each pressure level the optimization is solved numerically.
Since it is reasonable to assume that nearby pressure levels are characterized
by a similar \textbf{$\boldsymbol{\theta}$}, the initial value for
$\mathbf{\zeta}$ is set to zero for all $j=1,\ldots,M$. To avoid local minima,
the optimization for \textbf{$\boldsymbol{\theta}_{1}$} is repeated $100$
times with randomly perturbed initial values and \textbf{$\boldsymbol{\hat
{\theta}}_{1}$} is taken as the optimum of these $100$ solutions. The diagonal
variance covariance matrix $\Sigma_{\zeta}$, which acts as a smoothing factor
has been obtained by a preliminary not regularized estimation run.

\begin{figure}[t]
\centering\includegraphics[width=10cm]{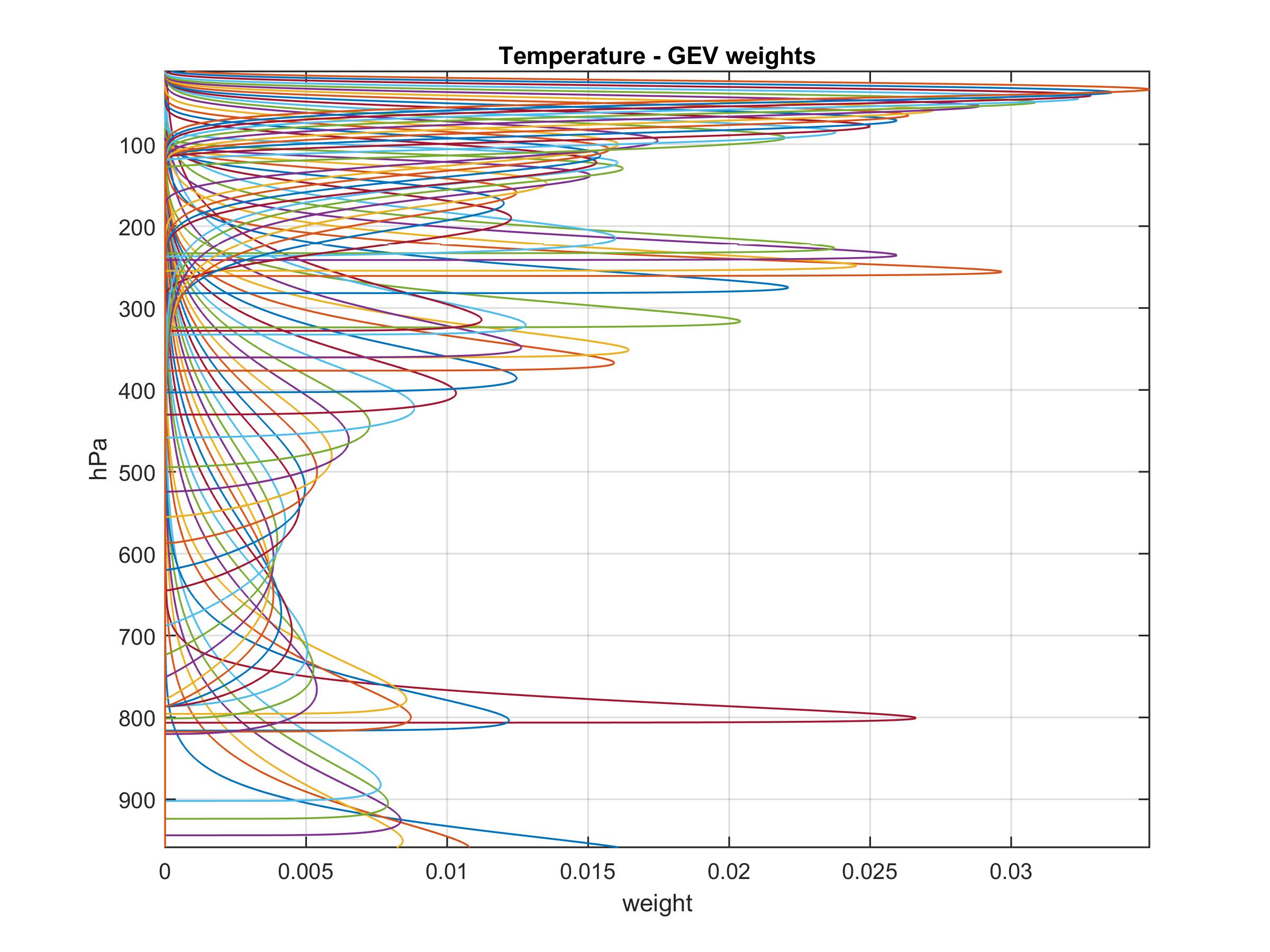}
\caption{Temperature. Weight functions given by GEV pdf's for IASI pressure
levels.}%
\label{fig:weights_temperature}%
\end{figure}

\begin{figure}[t]
\centering\includegraphics[width=10cm]{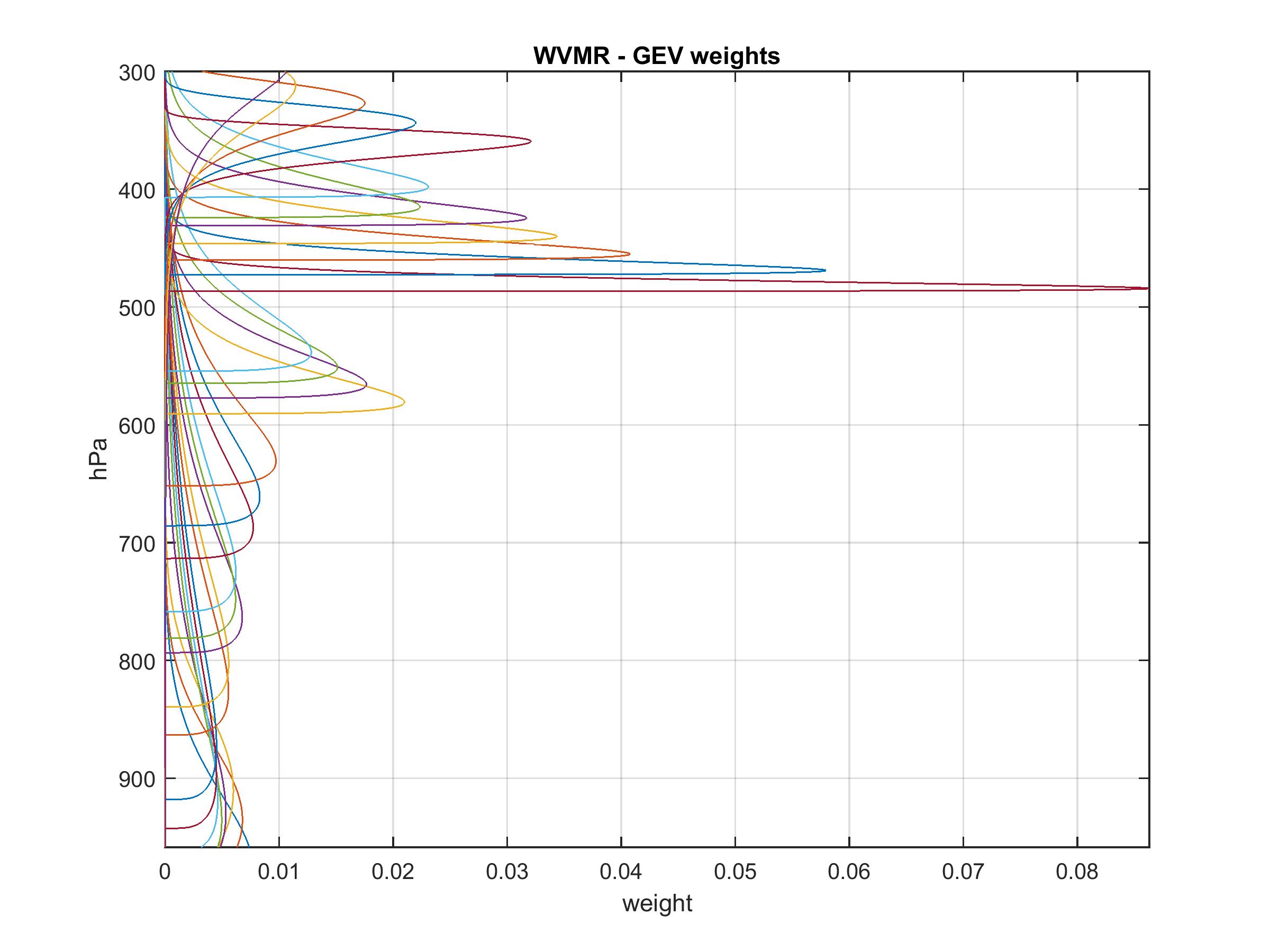} \caption{WVMR. Weight
functions given by GEV pdf's for IASI pressure levels.}%
\label{fig:weights_wvmr}%
\end{figure}

In order to illustrate the results, Figures \ref{fig:weights_temperature} and
\ref{fig:weights_wvmr} show GEV pdf's $w\left(  \cdot,\boldsymbol{\theta}%
_{j},p_{j}\right)  $ related to IASI pressure levels $p_{1},...,p_{M}$ for
temperature and WVMR. Each function essentially mimics how IASI sounds the
atmosphere at each specific pressure level with a peak near the corresponding
IASI\ level. The smaller the function width at pressure level $p_{j}$, the
better the IASI retrieval describes the ECV at that level.

Note that, in Figure \ref{fig:weights_temperature}, the pdf dispersion tends
to decrease at upper altitudes, especially above $50$ $hPa$. This is mainly
due to the non linearity of the pressure scale. For instance, a pressure
difference of $10$ $hPa$ at $20$ $hPa$ corresponds to an altitude difference
of around $4$ $km$, while the same pressure difference at $1000$ $hPa$
corresponds to an altitude difference of only $0.08$ $km$. Moreover in Figure
$\ref{fig:weights_wvmr}$ weight functions near $300$ $hPa$ are clearly
affected by a border effect and should be interpreted with caution.

\qquad

\begin{figure}[t]
\centering\includegraphics[width=10cm]{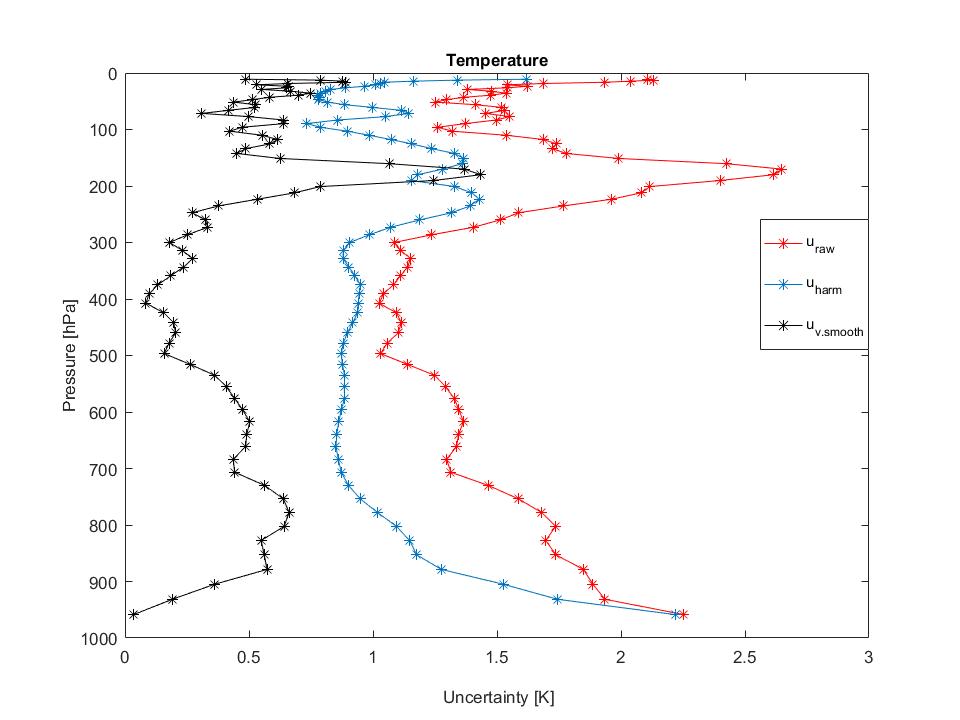}\caption{Temperature.
Vertical smoothing uncertainty: $u_{RI.vsmooth}$, black dotted line; it is the
uncertainty due to difference in smoothing between RAOB and IASI, profile
average is $0.501$ $K$. Harmonized mismatch uncertainty: $u_{RI.harm}$, cyan
dashed line; it is the uncertainty due to mismatch after adjusting for
difference in vertical smoothing, profile average is $1.052$ $K$. Unadjusted
uncertainty: $u_{RI.raw}$, red solid line; it is the total uncertainty between
interpolated RAOB and IASI, profile average is $1.553$ $K$. Formulas given in
Section \ref{sec:vertical.smoothing.uncertainty}.}%
\label{fig:u.temperature}%
\end{figure}

\begin{figure}[t]
\centering\includegraphics[width=10cm]{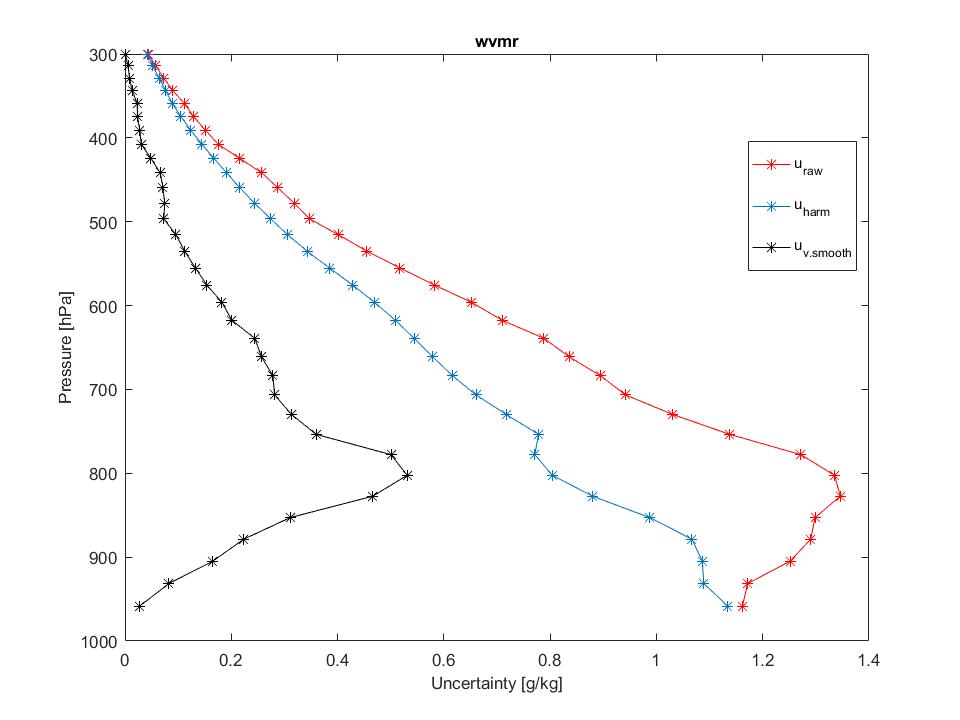}\caption{WVMR. Vertical
smoothing uncertainty: $u_{RI.vsmooth}$, black dotted line; it is the
uncertainty due to difference in smoothing between RAOB and IASI, profile
average is $0.1632$ $g/kg$. Harmonized mismatch uncertainty: $u_{RI.harm}$,
cyan dashed line; it is the uncertainty due to mismatch after adjusting for
difference in vertical smoothing, profile average is $0.4833$ $g/kg$.
Unadjusted uncertainty: $u_{raw}$, red solid line; it is the raw uncertainty
between interpolated RAOB and IASI, profile average is $0.6465$ $g/kg$.
Formulas given in Section \ref{sec:vertical.smoothing.uncertainty}.}%
\label{fig:u.wvmr}%
\end{figure}

After harmonizing RAOB to IASI, the adjusted uncertainty of Equation $\left(
\ref{eq:u.harmonized}\right)  $ is computed and the related vertical smoothing
uncertainty decomposition of Equation $\left(  \ref{eq:VS.u.decomposition}%
\right)  $, is reported in Figures \ref{fig:u.temperature} and
\ref{fig:u.wvmr}. The IASI-RAOB comparison is dominated by the smoothness of
the IASI retrieval and its reduced capability to catch strong vertical
gradients with respect to the RAOB profiles, though their sparseness. In the
boundary layer (BL) below about 900 hPa, where significant inversion in the
temperature profiles may occur, the hamonization does not strongly reduce the
raw uncertainty but above, up to 700 hPa, the reduction becomes more
significant. In the upper troposphere/lower stratosphere the strong gradients
at the tropopause increase the raw uncertainty and the harmonization strongly
reduces the difference between IASI and RAOB. It is worth to remind that the
values calculated above 100 hPa are affected by the size of the sampling which
is more limited that at higher pressure levels. For WVMR, results similar to
temperature are observed in the BL. The increase of the smoothness uncertainty
at 800 hPa is likely linked to the transition from wetter to drier air
occurring at the top of BL not always caught in the RAOB data. The benefit of
the harmonization decreases with the altitude proportionally with the decrease
of the water vapour variability in the atmosphere. This is clearly visible
from the difference between the raw and harmonized uncertainties which reduces
with the height.

\section{Discussion and conclusion\label{sec:Conclusions}}

This paper discussed the comparison of IASI and RAOB temperature and humidity
with a focus on vertical smoothing. Since the IASI averaging kernels have been
considered unknown, a weighting function mimicking the weights of the
averaging kernel has been estimated on data. To do this RAOB data have been
transformed in functional data and the related uncertainty has been assessed
by a comparison with the reference measurements for radiosonde given by GRUAN
data, Lindenberg. Hence it can be considered as a first substantial step in
the direction of Calbet et al. 2017 "To fully characterize the comparison, a
method to estimate the collocation uncertainty would be desirable. This method
should not depend on the data being used for the study and should be
independent from them".

Thanks to this approach, it has been found that the uncertainty of vertical
smoothing mismatch averaged along the profile is $0.50$ $K$ for temperature
and $0.16$ g/kg for water vapour mixing ratio. Moreover, the uncertainty
related to RAOB vertical sparseness, averaged along the profile is $0.29$ $K$
for temperature and $0.13$ g/kg for water vapour mixing ratio.

From the methodological point of view, it has been shown that the estimates
are obtained by the maximum likelihood method, taking into consideration also
the measurement uncertainties where available.

\subsection{Further developments}

Further aspects may be added to the presente analysis of satellite vs
radiosonde comparison. For instance, the distance between the satellite line
of sight and the radiosonde position has not been considered. In fact this
issue will be addressed in a forthcoming paper using isotonic regression
(Mayer, $2013$).

Vertical correlation has not been considered explicitly here and, in this
sense, the results could be suboptimal. In fact, IASI measurements are known
to have a limited number of degrees of freedom. In our approach at least a
part of IASI vertical correlation is implicitly handled by the random walk
dynamics of GEV pdf parameters in Section \ref{sec:IASI.RAOB.likelihood}.
Considering radiosonde, sources of vertical correlation arise both from short
range smoothing algorithms, used to avoid measurement outliers, and by
pre-launch calibration errors. Although a good part of these problems is
automatically handled by the functional data approach used here, further
research could point out new solutions.

A further insight into vertical smoothing could benefit from the comparison of
this proposal with the IASI "true averaging kernel" at least in some cases.
Nonetheless, we remark that the approach of this paper can be used even in
absence of averaging kernels, which is quite relevant especially for
historical records.

\section*{Acknowledgements}

This research is partially funded by GAIA-CLIM, the project funded from the
European Union's Horizon 2020 research and innovation programme under grant
agreement No 640276.

We are very thankful to NOAA and in particular to Tony Reale for providing the
radiosonde-satellite collocations.


\begin{thebibliography}{99}                                                                                               %


\bibitem {}Berrisford P, Dee DP, Fielding K, Fuentes M, K\aa llberg P,
Kobayashi S, Uppala SM. $\left(  2009\right)  $ `The ERA-Interim Archive'. ERA
Report Series, No. 1. ECMWF: Reading, UK.

\bibitem {}Bojinski, S., Verstraete, M., Peterson, T. C., Richter, C.,
Simmons, A., and Zemp, M. $\left(  2014\right)  $ The concept of Essential
Climate Variables in support of climate research, applications, and policy,
\emph{B. Am. Me- teorol. Soc.}, 95, 1431--1443, doi:10.1175/bams-d-13-00047.1.

\bibitem {}Calbet, X., Peinado-Galan, N., Ripodas, P., Trent, T., Dirksen, R.,
and Sommer, M. $\left(  2017\right)  $ Consistency between GRUAN sondes,
LBLRTM and IASI. \emph{Atmos. Meas. Tech}., 10(6), 2323-2335. https://doi.org/10.5194/amt-10-2323-2017.

\bibitem {}Dirksen, R. J., Sommer, M., Immler, F. J., Hurst, D. F., Kivi, R.,
and V\"{o}mel, H. $\left(  2014\right)  $ Reference quality upper-air
measurements: GRUAN data processing for the Vaisala RS92 radiosonde.
\emph{Atmos. Meas. Tech}., 7, 4463-4490, https://doi.org/10.5194/amt-7-4463-2014.

\bibitem {}Fahrmeir L., Kneib T., Lang S. \& Marx B. $\left(  2013\right)  $
Regression - Models, Methods and Applications. Springer. Berlin.

\bibitem {}Fass\`{o}, A., Ignaccolo, R., Madonna, F., Demoz, B. and
Franco-Villoria M. $\left(  2014\right)  $ Statistical modelling of
collocation uncertainty in atmospheric thermodynamic profiles. \emph{Atmos.
Meas. Tech.} 7, 1803--1816.

\bibitem {}Ignaccolo, R., Franco-Villoria, M., Fass\`{o}, A. $\left(
2015\right)  $ Modelling collocation uncertainty of 3D atmospheric profiles.
\emph{Stochastic Environmental Research and Risk Assessment} 29 (2), 417-429.

\bibitem {}Immler, F. J., Dykema, J., Gardiner, T., Whiteman, D. N., Thorne,
P. W., and V\"{o}mel, H. $\left(  2010\right)  $ Reference Quality Upper-Air
Measurements: guidance for developing GRUAN data products, Atmos. Meas. Tech.,
3, 1217--1231, doi:10.5194/amt-3-1217-2010.

\bibitem {}Kotz, S., Nadarajah S. $\left(  2000\right)  $ \emph{Extreme Value
Distributions: Theory and Applications}. Imperial College Press, London, UK.

\bibitem {}Kursinski, E. R., and G. A. Hajj, $\left(  2001\right)  $ A
comparison of water vapor derived from GPS occultations and global weather
analyses, \emph{J. Geophys. Res}., 106, 1113--1138.

\bibitem {}Lambert, J.C. and Vandenbussche, S. $\left(  2011\right)  $
Multi-dimensional characterisation of remotely sensed data. EC FP6 GEOmon
Technical Note D4.2.1. GEOmon TN-IASB-OBSOP, BIRA-IASB.

\bibitem {}Meyer, M.C. $\left(  2013\right)  $ A Simple New Algorithm for
Quadratic Programming with Applications in Statistics, \emph{Communications in
Statistics}, 42(5), 1126-1139.

\bibitem {}Pappalardo, G., Wandinger, U., Lucia, M., Hiebsch, A. Mattis, I.,
et al. (2010), EARLINET correlative measurements for CALIPSO: First
intercomparison results, J. Geophys. Res., 115, D00H19, doi:10.1029/2009JD012147.

\bibitem {}Pougatchev, N., August, T., Calbet, X., Hultberg, T., Oduleye, O.,
Schl\"{u}ssel, P., Stiller, B., Germain, K. St., and Bingham, G. $\left(
2014\right)  $ IASI temperature and water vapor retrievals -- error assessment
and validation, \emph{Atmos. Chem. Phys.}, 9, 6453--6458, doi:10.5194/acp- 9-6453-2009.

\bibitem {}Ravegnani, F., Reburn, W. J., Redaelli, G., Remedios, J. J.,
Sembhi, H., Smale, D., Steck, T., Taddei, A., Varotsos, C., Vigouroux, C.,
Waterfall, A., Wet- zel, G., and Wood, S. $\left(  2007\right)  $ Geophysical
validation of MIPAS- ENVISAT operational ozone data, \emph{Atmos. Chem.
Phys.}, 7, 4807-- 4867, doi:10.5194/acp-7-4807-2007.

\bibitem {}Ramsay, J., Silverman, B. $\left(  2012\right)  $ Applied
Functional Data Analysis. Methods and Case Studies. Springer.

\bibitem {}Reinsch C. $\left(  1971\right)  $ "Smoothing by spline functions
II", Numerische. Mathematik. 16, 451-454.

\bibitem {}Ridolfi, M., Blum, U., Carli, B., Catoire, V., Ceccherini, et al.
$\left(  2007\right)  $ Geophysical validation of temperature retrieved by the
ESA processor from MIPAS/ENVISAT atmospheric limb- emission measurements,
\emph{Atmos. Chem. Phys.}, 7, 4459--4487, doi:10.5194/acp-7-4459-2007.

\bibitem {}Seidel, D. J., B. Sun, M. Pettey, and A. Reale $\left(
2011\right)  $, Global radiosonde balloon drift statistics, J. Geophys. Res.,
116, D07102, doi:10.1029/2010JD014891

\bibitem {}Sun, B., Reale, A., Tilley, F., Pettey, M., Nalli, N. R., Barnet,
C. D. $\left(  2017\right)  $ Assessment of NUCAPS S-NPP CrIS/ATMS sounding
products using reference and conventional radiosonde observations. \emph{IEEE
Journal of Selected Topics in Applied Earth Obsrevations and Remote Sensing}.
10:6, pp 2499-2509. DOI: 10.1109/JSTARS.2017.2670504.

\bibitem {}Tobin, D. C., Revercomb, H. E., Knuteson, R. O., Lesht, B. M.,
Strow, L. L., Hannon, S. E., Feltz, W. F., Moy, L. A., Fetzer, E. J., and
Cress, T. S. $\left(  2006\right)  $ Atmospheric Radiation Measurement site
at- mospheric state best estimates for Atmospheric Infrared Sounder
temperature and water vapor retrieval validation, \emph{J. Geophys. Res.},
111, D09S14, doi:10.1029/2005JD006103.

\bibitem {}Verhoelst, T., Granville, J., Hendrick, F., K\"{o}hler, U., Lerot,
C., Pommereau, J.-P., Redondas, A., Van Roozendael, M., and Lambert, J.-C.
$\left(  2015\right)  $ Metrology of ground-based satellite validation:
co-location mismatch and smoothing issues of total ozone comparisons,
\emph{Atmos. Meas. Tech.}, 8, 5039-5062, doi:10.5194/amt-8-5039-2015.

\bibitem {}Maddy, E.S., and Barnet, C.D. $\left(  2008\right)  $ Vertical
resolution estimates in version 5 of AIRS operational retrievals. \emph{IEEE
Transactions on Geoscience and Remote Sensing}. 46, pp. 2375-2384.
\end{thebibliography}
\end{document}